\documentclass[onecolumn,11pt]{IEEEtran}

\usepackage{amsmath}
\usepackage{amsfonts,amssymb}
\usepackage{multirow}
  \usepackage[pdftex]{graphicx}

\def\bv{\mathbf{v}}

\def\bV{\mathbf{V}}
\def\bX{\mathbf{X}}
\def\bY{\mathbf{Y}}

\def\bx{\mathbf{x}}
\def\by{\mathbf{y}}

\def\EE{\mathbb{E}}

\def\FF{\mathbb{F}}
\def\CC{\mathbb{C}}
\def\RR{\mathbb{R}}

\def\argmax{\mathop{{\rm argmax}}\limits}
\def\mix{\mathop{{\rm mix}}\nolimits}

\def\im{\mathop{{\rm Im}}\nolimits}

\newtheorem{theorem}{Theorem}
\newtheorem{lemma}{Lemma}

\def\Label#1{\label{#1}\ [\ \text{#1}\ ]\ }
\def\Label{\label}

\newenvironment{proofof}[1]{\vspace*{5mm} \par \noindent
         \quad{\it Proof of #1:\hspace{2mm}}}{\endIEEEproof
\hfill$\Box$ 
}

\begin{document}
\title{Secure Computation-and-Forward with Linear Codes}

\author{Masahito~Hayashi
\thanks{The first author is with the Graduate School of Mathematics, Nagoya University, Japan. He is also with the Center for
Quantum Technologies, National University of Singapore, Singapore, e-mail:masahito@math.nagoya-u.ac.jp},
                        Tadashi Wadayama
\thanks{The second author is with 
Department of Computer Science, Faculty of Engineering, Nagoya Institute of Technology, Japan,
e-mail: wadayama@nitech.ac.jp.},
and 
\'{A}ngeles Vazquez-Castro
\thanks{The third author is with 
Department of Telecommunications and Systems Engineering,
Autonomous University of Barcelona
e-mail: 
angeles.vazquez@uab.cat}
}
\maketitle

\begin{abstract}
We discuss secure transmission via an untrusted relay when 
we have a multiple access phase from two nodes to the relay
and broadcast phase from the relay to the two nodes.
To realize the security, 
we construct a code that securely transmits the modulo sum of the messages of two nodes
via a multiple access channel.
In this code, the relay cannot obtain any information for the message of each node,
and can decode only the messages of the two nodes.
Our code is constructed by simple combination of an existing liner code and universal2 hash function.
\end{abstract}

\begin{IEEEkeywords} 
secrecy analysis,
secure communication,
interference,
relay,
multiple access channel,
computation and forward
\end{IEEEkeywords}

\section{Introduction}
The two-way relay network model is a cooperative communication network that consists of two nodes $1$ and $2$ that want to communicate to each other but there is no direct link between them \cite{ZLWL,RW}. 
The intermediate relay node $R$ assists the communication between the two nodes. Communication takes place in two phases, a multiple access (MAC) phase and a broadcasting phase as Fig. \ref{FF1}. 
Transmissions are assumed perfectly synchronised and the communication in the MAC and broadcasts phases are orthogonal. 
The relay node decodes the sum of the messages from the two nodes in the uplink,
and broadcasts it to the two nodes in the downlink.
However, a part of information from each message is leaked to the relay node.
When the relay node is untrusted, it is needed to keep the secrecy of both message from the relay node.
That is, the following task is required.
When Nodes $1$ and $2$ have the messages $M_1$ and $M_2$, 
the relay node decodes the module sum $M_1+M_2$ without obtaining information for $M_1$ and $M_2$. 
We call this task secure computation-and-forward.
The preceding papers \cite{Ren,He1,He2,Vatedka,Zewail} discussed 
secure computation-and-forward by using the lattice code with computation-and-forward.
However, the lattice code has large cost for its implementation 
because the number of constellation points increases when the size of code increase.
Even if multilevel implementations have been proposed \cite{HY},
it is better to employ a linear code with fixed constellation points.
Moreover, it is desirable that the employed code has encoding and decoding with small computational complexity.

\begin{figure}[t]
\begin{center}
  \includegraphics[width=0.7\linewidth]{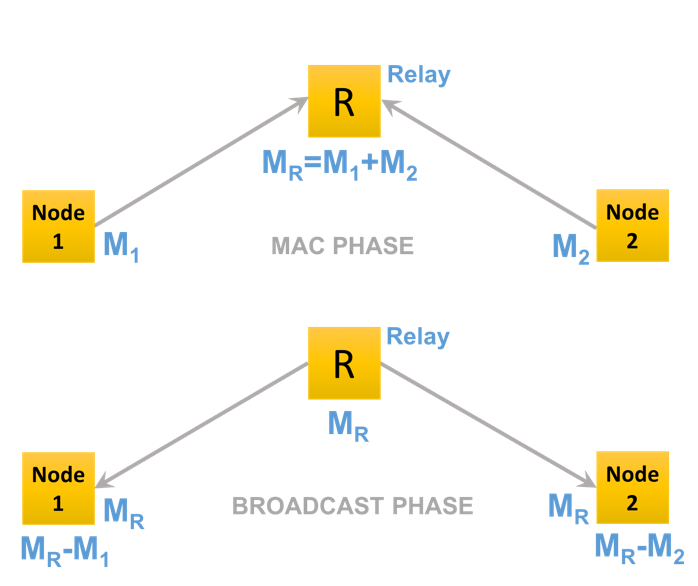}
  \end{center}
\caption{MAC phase and broadcast phase.}
\Label{FF1}
\end{figure}

For this aim, as a typical scenario, we focus on a multiple access channel 
and address use of the channel $n$ times
when two users' input alphabets are given as $\FF_q$
and their constellation points are fixed.
Then, we fix a sequence of general linear codes in $\FF_q^n$.
Similar to \cite{Expo,VH}, using the sequence of linear codes and attaching universal2 hash function,
we construct a sequence of codes with strong secrecy for the untrusted relay.
For a practical use, we can choose error correcting codes with efficient decoder, e.g.,
LDPC codes, as the general linear codes.
Then, we derive 
the amount of the leaked information in the finite-length setting, which is required to guarantee the secrecy in an implemented system.
Recently, Takabe et al. \cite{Takabe} addressed 
this kind of Gaussian multiple access channel with $\FF_2$ when the sum of the messages of both nodes is decoded.
Then, using density evolution method,
they derived the threshold of standard deviation of noises of 
spatially coupled LDPC codes with belief propagation decoding, which implies 
the threshold of decodable rate.
Hence, it is useful to apply these error correcting codes to our secure code construction.
In this paper, we derive the asymptotic transmission rate of this practical code.
Then, we apply our finite-length secrecy evaluation to this practical code.

As another application of secure computation-and-forward,
we consider 
butterfly network coding, which is a coding method that efficiently transmits the information 
in the crossing way as Fig. \ref{F8}.
However, when the secrecy of the message is required,
this conventional butterfly network coding has the following problem.
The intermediate node $V_2$ 
can obtain the information of the messages.
Also, the receiver nodes $V_5$ and $V_6$ can obtain the information of the other message, respectively.
Although secure network coding is known,
it cannot realize this kind of secrecy in the butterfly network.
When we apply secure computation-and-forward to 
the communications to nodes $V_2, V_5,$ and $V_6$,
the desired secrecy is realized.

\begin{figure}[h]
\begin{center}
\includegraphics[scale=0.7, angle=-90]{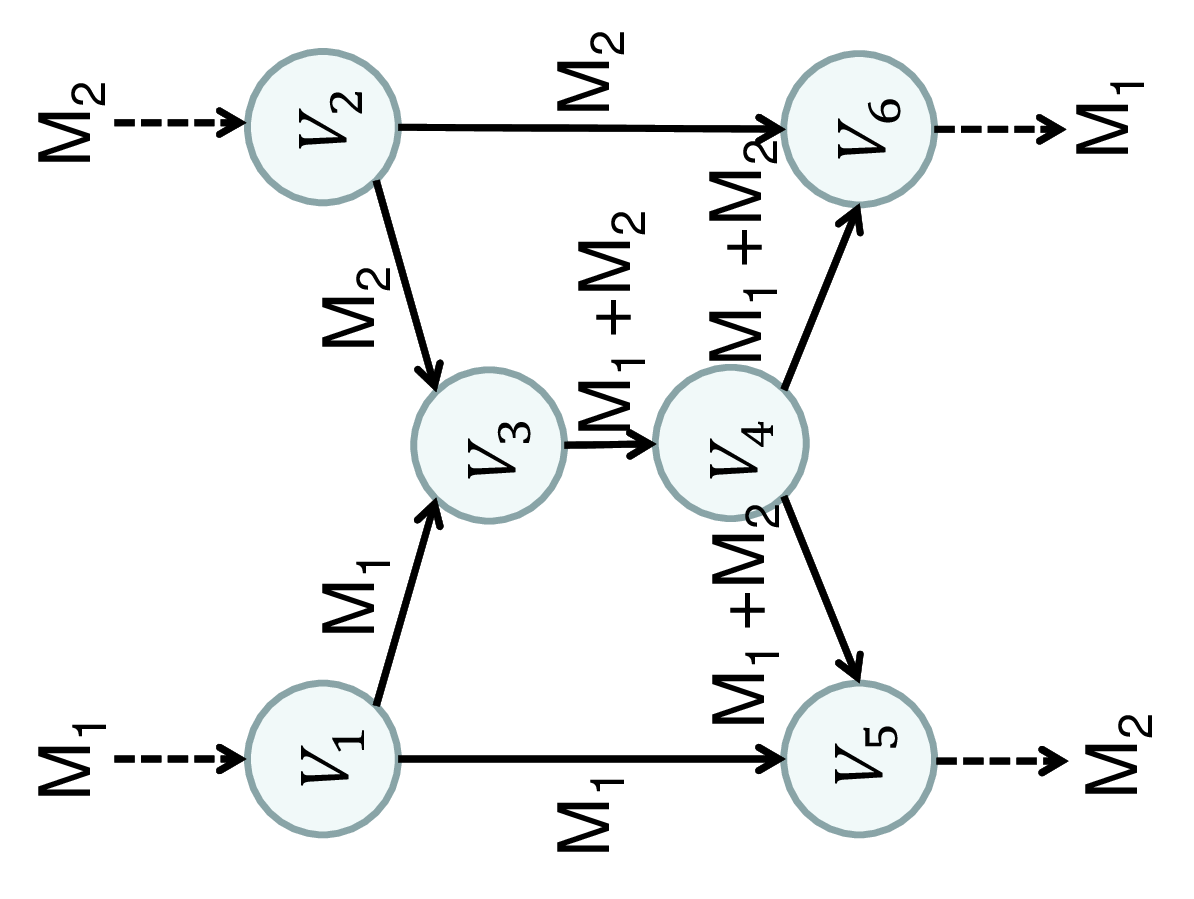}
  \end{center}
\caption{Butterfly network coding.}
\Label{F8}
\end{figure}

The remaining of this paper is organized as follows.
Using a linear code for computation-and-forward,
Section \ref{S2} constructs our secure code that has no information leakage to the relay node.
Section \ref{S3} gives security analyses with the finite-length setting.
Section \ref{S5} numerically evaluates the asymptotic achievable rate 
in the cases of random coding and spatial coupling LDPC code with BPSK scheme.

\section{Code Construction}\Label{S2}
When we use the channel $n$ times,
we discuss a secure protocol to exchange their messages $M_1$ and $M_2$
without information leakage to the relay $R$.
In the MAC phase protocol, 
given an arbitrary map $\sigma$ from $\FF_q$ to $\RR$ or $\CC$,
we assume the following MAC channel $W$
\begin{align}
\textbf{Y}_R=h_1 \sigma (\bX_1) + h_2 \sigma (\bX_2) +\textbf{Z}_R,
\Label{23-1}
\end{align}
where
$h_1, h_2 \in \RR$ or $\CC$ are the channel fading coefficients 
and 
$\textbf{Z}_R \sim \mathcal{N}(0, N_0\textbf{I}_n)$ is a vector of jointly Gaussian real random variables. 
Here, 
$\bY_R $ is an $n$-dimensional real or complex value,
and $\bX_1 $ and $ \bX_2 $ are $n$-dimensional vectors of $\FF_q$.

As a typical example, we often employ the BPSK scheme, i..e, $q=2$.
Then, we fix a map $\sigma$ from $x\in \FF_2$ to $\RR$ as
$(-1)^x $.
Then, our multiple input channel channel is given as the map
$W: (x_1,x_2)\mapsto \phi_{h_1 \sigma(x_1)+h_2 \sigma(x_2) ,N_0} $,
where $\phi_{a,N_0} $ is the Gaussian distribution with average $a$ and variance $N_0$.

Now, we assume that node $i$ encodes the information $\bV_i \in \FF_q^{k_n}$ 
instead of $M_i$ and the relay $R$ recovers $\bV_1+\bV_2 \in \FF_q^{k_n}$. 
In this case, both nodes often employ the same linear map $G: \FF_q^{k_n} \to \FF_q^n $ with rank $k$ as an encoder and 
relay $R$ employs a decoder $D$, which is a map from $\RR^n$ or $\CC^n$ to 
$\FF_q^{k_n}$. 
Here, relay $R$ is assumed to know the coefficients $h_1$, $h_2$, the map $\sigma$, and $N_0$.

Now, we discuss the scheme with shift vectors $ e_1, e_2 \in \FF_q^n$ as follows.
The encoder $\Phi_{G,E_1,1}^n$ of node 1 maps $\bV_1 \in \FF_q^{k_n}$
to the element $ G(\bV_1)+e_1$ of the alphabet,
and the encoder $\Phi_{G, e_2,2}^n$ of the node 2 maps $\bV_2 \in \FF_q^{k_n}$
to the element $ G(\bV_2)+e_2$ of the alphabet.
As decoding process, the relay $R$ obtains 
$\bV_R:=D(\bY_R -h_1\sigma (e_1)-h_2\sigma (e_2))$.

In the broadcast phase protocol,
the relay $R$ sends the information 
$\bV_R$ to nodes 1 and 2, which can be achieved by a conventional channel coding.
Since node 1 has the information $\bV_1$, 
node 1 recovers the information $\bV_2$ as $\bV_R-  \bV_1 $.
Similarly, node 2 recovers the information $\bV_1$.

Our interest is information leakage to the relay $R$.
Now, we discuss a secure protocol to exchange their messages $M_1$ and $M_2$
without information leakage to $R$.
To discuss this problem, we discuss a slightly different protocol.
When $\bV_1$ and $\bV_2$ are uniform random number, 
we have the relations $I(\bV_R;\bV_i)=0$ for $i=1,2$, i.e.,
$\bV_R$ is independent of $\bV_i$ with $i=1,2$.
However, the relay $R$ obtains the information
\begin{align}
\textbf{Y}_R=h_1 \sigma (G (\bV_1)+e_1) +h_2 \sigma (G (\bV_2)+e_2)+\textbf{Z}_R,
\end{align}
which is more informative than $\bV_R$.
Further, the variable $\textbf{Y}_R$ has correlation with $\bV_i$ for $i=1,2$.
This information leakage can be removed when
nodes 1 and 2 apply a linear hash function $F:\FF_q^{k_n}\rightarrow\FF_q^{k_n-\bar{k}_n}$ whose rank is $k_n-\bar{k}_n$.

Now, we prepare the auxiliary random variable $L_i \in \FF_q^{\bar{k}_n}$ for $i=1,2$.
We choose linear functions $F_1:\FF_q^{k_n-\bar{k}_n} \to \FF_q^{k_n}$ and 
$F_2:\FF_q^{\bar{k}_n} \to \FF_q^{k_n}$ such that
$F \circ F_1 $ is the identity map on $\FF_q^{k_n-\bar{k}_n}$ and 
the image of the map 
$(m_1,l_2)\in \FF_q^{k_n-\bar{k}_n} \times \FF_q^{\bar{k}_n} \mapsto 
F_1(m_1)+F_2(l_2)$ is $\FF_q^{k_n}$.
Then, the encoders is given as
$\Phi_{G,e_1,1}^n(F_1(M_1)+ F_2(L_1)) $
and
$\Phi_{G,e_2,2}^n(F_1(M_2)+ F_2(l_2)) $.
That is, the random variable $\bV_i$ is given as 
$F_1(M_i)+ F_2(L_i)$.
The decoder is given as 
$F (D(\bY_R -h_1\sigma (e_1)-h_2\sigma (e_2)))$.
The relay $R$ broadcasts it.
Then, we denote the above protocol with a linear map $G$ and shift vectors $ e_1, e_2$ of block length $n$ by $\Phi_{G ,e_1,e_2}^n$.
In summary, the encoding and decoding processes are illustrated as Fig. \ref{FT}.

\begin{figure}[h]
\begin{center}
  \includegraphics[width=0.9\linewidth]{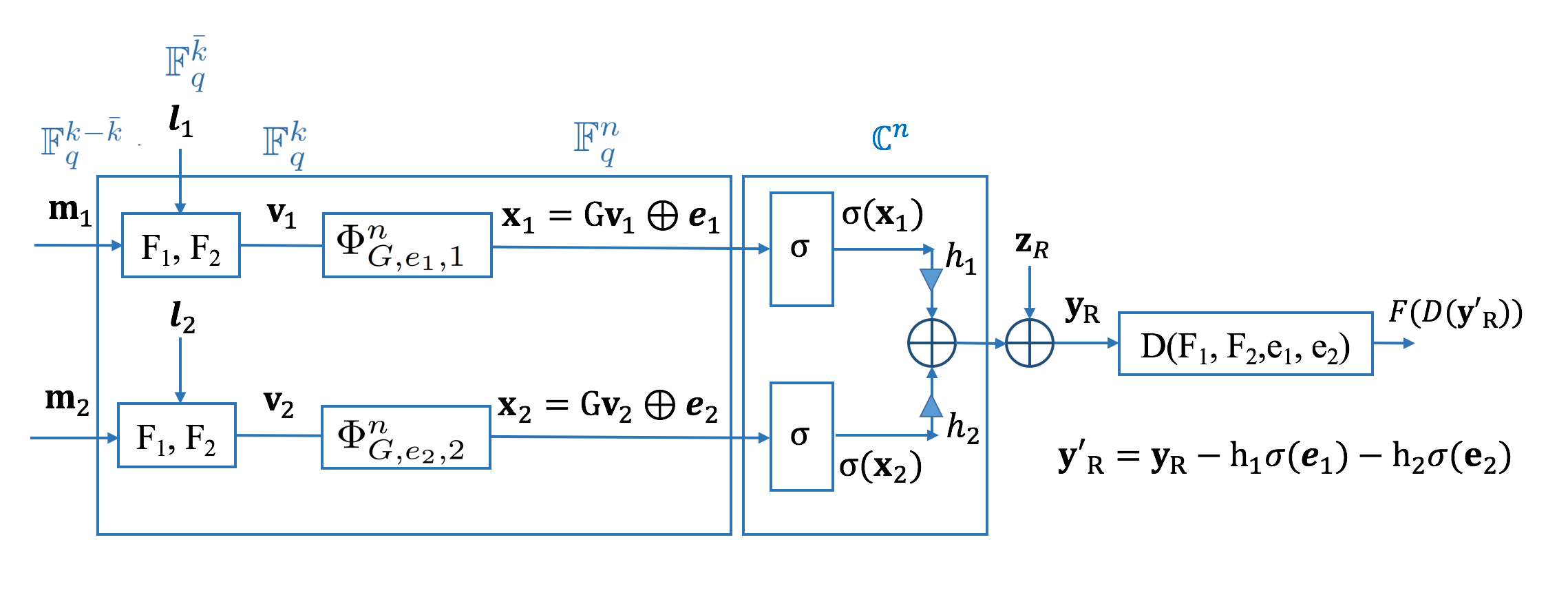}
  \end{center}
\caption{Encoding and decoding process.}
\Label{FT}
\end{figure}

\section{Secrecy Analysis with Transmission Rate}\Label{S3}
In this section, we derive a finite-length bound for leaked information when 
${\cal X}=\FF_q$.
To discuss the information leakage for $M_i$, we introduce the security criterion for $i=1,2$
\begin{align}
& d(\Phi_{G,e_1,e_2}^n)_i \nonumber \\
:=&
\| P_{M_i \bY| E_1=e_1, E_2=e_2 ,n}-  
P_{\bY| E_1=e_1, E_2=e_2 ,n}\times P_{M_i,\mix}\|_1 \nonumber \\
=& \sum_{m_i} P_{M_i} (m_i)
\int_{\CC} \Big| p_{\bY|M_i, E_1=e_1, E_2=e_2,n }(\by|m_i)\nonumber \\
&\quad - p_{\bY| E_1=e_1, E_2=e_2,n }(\by) \Big| d \by,
\end{align}
where $P_{M_i,\mix}$ expresses the uniform distribution for $M_i$.

In the following, for security analysis,
priorly, the shift vectors $e_1$ and $e_2$ are chosen randomly.
So, they are treated as random variables, and are denoted by $E_1$ and $E_2$.
We consider the case when $G$ is chosen as a code with efficient decoder.


For finite-length analysis,
we prepare other notations and information quantities used in this paper.
Given a joint distribution channel $P_{Y,Z_1, Z_2}$ over the product system of 
a finite discrete set ${\cal Z}_1\times {\cal Z}_2$ and a continuous set ${\cal Y}$,
we denote the conditional probability density function of 
$P_{Y|Z_1,Z_2}$ by $p_{Y|Z_1,Z_2}(y|z_1,z_2) $.
Then, we define the conditional distribution $P_{Y| Z_1}$ 
over a continuous set ${\cal Y}$ conditioned in the discrete set ${\cal Z}_1$
by the conditional probability density function
$p_{Y|Z_1}(y|z_1):= \sum_{z_2\in {\cal Z}_2} P_{Z_2}(z_2) p_{Y|Z_1,Z_2}(y|z_1,z_2)$.
Then, we define the Renyi conditional mutual information
$I_{1+s}^{\downarrow}( Y;Z_1|Z_2)$
\begin{align}
& \frac{s}{1+s} I_{1+s}^{\downarrow}( Y;Z_1|Z_2)\nonumber \\
:=&
\log 
\sum_{z_2}
P_{Z_2}(z_2)
\int_{{\cal Y}}  
\Big(
\sum_{z_1}
P_{Z_1|Z_2}(z_1|z_2)
p_{Y|Z_1,Z_2}(y|z_1,z_2)^{1+s}
\Big)^{\frac{1}{1+s}}
dy 
\end{align}
for $s>0$.
Since $\lim_{s \to 0}s I_{1+s}^{\downarrow}( Y;Z_1|Z_2)=0$,
taking the limit $s \to 0$, we have
\begin{align}
\lim_{s \to 0}\frac{s I_{1+s}^{\downarrow}( Y;Z_1|Z_2)}{s} =
I( Y;Z_1|Z_2),
\end{align}
where 
$I( Y;Z_1|Z_2)$ expresses the conditional mutual information. 
The concavity of the function $x \mapsto x^{\frac{1}{1+s}}$ yields
\begin{align}
e^{\frac{s}{1+s} I_{1+s}^{\downarrow}( Y;Z_1|Z_2,Z_3)}
\le e^{\frac{s}{1+s} I_{1+s}^{\downarrow}( Y;Z_1,Z_2|Z_3)}.\Label{HY2}
\end{align}

Given a channel $P_{Y|Z_1, Z_2,Z_3}$ from the finite discrete set ${\cal Z}_1\times {\cal Z}_2\times {\cal Z}_3$
to a continuous set ${\cal Y}$,
when the random variables $Z_1,Z_2,Z_3$ are generated subject to the uniform distributions,
we have a joint distribution among $Y,Z_1,Z_2$.
In this case, we denote 
the mutual information $I(X_1;Y)[P_{Y|Z_1,Z_2,Z_3}]$.
This rule is applied to 
the R\'{e}nyi conditional mutual information and the conditional mutual information as
 $I_{1+s}^{\downarrow}( Y;Z_1|Z_2)[P_{Y|Z_1,Z_2,Z_3}]$ and $I( Y;Z_1|Z_2)[P_{Y|Z_1,Z_2,Z_3}]$, respectively.
In the following, we use this notation to the channel
$W$ defined by 
\begin{align}
Y=h_1 \sigma (X_1) + h_2 \sigma (X_2) +{Z}_R,
\Label{23-1M}
\end{align}
where $Z_R$ is the Gaussian variable with average $0$ and the variance $N_0$ on
$\mathbb{R}$ or
$\mathbb{C}$.
Here, the choice of random variables $Z_1,Z_2,$ and $Z_3$ depends on the context.

\begin{theorem}\Label{TTA}
Given a map $G=g$, using 
$B_{i,n,s,1}:= 3 q^{s (n-k_n-\bar{k}_n)} e^{s n I_{\frac{1}{1-s}}^{\downarrow}
(Y; X_i  )[W] }$, 
we have 
\begin{align}
\EE_{E_1,E_2} d(\Phi_{g,E_1,E_2}^n)_i 
\le
\min_{s \in [0,\frac{1}{2}]} B_{i,n,s,1},
 \Label{LLLA}.
\end{align}
\end{theorem}

To improve the bound \eqref{LLLA},
we focus on the ensemble of injective linear codes
$G: \FF_q^{k_n} \to \FF_q^{n}$.
We consider  the permutation-invariance for the ensemble as follows.
We say that the ensemble $G$ is permutation-invariant when 
${\rm Pr}( \bx \in \im G )=  {\rm Pr}( g(\bx) \in \im G )$
for any $\bx \in {\cal X}^n$ and any permutation $g$ among $\{1, \ldots, n\}$.
In addition, we often consider the following condition.
We say that the ensemble $G$ is universal 2
when the ensemble $G$ satisfies the condition
\begin{align}
{\rm Pr}\{ x \in \im G\}\le q^{k_n-n}\Label{GDE}
\end{align}
holds for any $x (\neq 0) \in \FF_q^n$ \cite{Carter,Krawczyk}.

Let $\vec{\lambda}$ be an integer-valued vector $(\lambda_{t})_{t \in \FF_q}$ such that
$\sum_{t \in \FF_q} \lambda_t=n$, $\lambda_0 \neq n$, and 
$\lambda_t \ge 0$.
We denote the set of such integer-valued vectors by ${\cal T}_n(\FF_q)$.
For a code $g$, we define
\begin{align}
N(\vec{\lambda}, g ):=| \{ \bx \in \im g | \vec{n}(\bx) = \vec{\lambda}\}|,
\Label{TYO4}
\end{align}
where
$n_{t}(\bx)$ expresses the number of $t$ in the vector $\bx$.
Then, using the above number, we define the value
\begin{align}
A&:=\max_{\vec{\lambda}} 
A(\vec{\lambda}),\Label{TYO}\\
A(\vec{\lambda}) &:=\max_{\vec{\lambda} (\neq \vec{0}_n)\in {\cal T}_n(\FF_q)} \EE_{G} \frac{N(\vec{\lambda},G)q^{n-k_n}}{ {n \choose \vec{\lambda}} },\Label{TYO2}
\end{align}
where ${n \choose \vec{\lambda}} $ expresses the multi-nomial combination,
and $\vec{0}_n$ expresses the vector satisfying that
$(\vec{0}_n)_0=n$ and $(\vec{0}_n)_t=0$ for $t (\neq 0)\in \FF_q$.
When the ensemble $G$ is universal 2, that is, the ensemble $G$ has no deviation, 
we have $A\le 1$.
Hence, $A$ expresses the degree of deviation.

\begin{theorem}\Label{TTY}
When the ensemble $G$ is permutation-invariant,
\begin{align}
\EE_{E_1,E_2,G} d(\Phi_{G,E_1,E_2}^n)_i 
\le
\min_{s \in [0,\frac{1}{2}]} B_{i,n,s,2} [A]
\Label{LLL},
\end{align}
where $B_{i,n,s,2}[A]$ is defined to be
\begin{align}
3 q^{-s (k_n+\bar{k}_n)}
e^{s n I_{\frac{1}{1-s}}^{\downarrow}(Y; X_1 ,X_2  )[W] } 
+3 A^s q^{-s \bar{k}_n} e^{s n I_{\frac{1}{1-s}}^{\downarrow}
(Y; X_i  )[W] }.
 \Label{LLL7}
\end{align}
\end{theorem}

Although Theorem \ref {TTY} assumes the permutation-invariance,
we do not need this condition because of the following reason.
We focus on a code $g$.
When the ensemble $G$ is given as the ensemble given by the application of the random permutation to the code $g$,
$\EE_{G} d(\Phi_{G,E_1,E_2}^n)_i 
= d(\Phi_{g,E_1,E_2}^n)_i $ because 
the amount of leaked information $d(\Phi_{g,E_1,E_2}^n)_i $ does not change
under the application of permutation.

As the comparison between \eqref{LLLA} and \eqref{LLL}, we have the following lemma, whose proof is given in Appendix \ref{AA}.

\begin{lemma}\Label{GUR}
We have
\begin{align}
2 B_{i,n,s,1}
\ge 
B_{i,n,s,2}[A]
 \Label{LLL6}.
\end{align}
That is,
the bound \eqref{LLL} is smaller than the twice of the bound \eqref{LLLA}.
\end{lemma}

In the following, we derive the achievable rates based on the upper bounds \eqref{LLLA} and \eqref{LLL}.
For this aim, we introduce the parameter $r_2$ for our sequence of code ensembles satisfying 
\begin{align}
\lim_{n\to \infty}\frac{\log A}{n}= \frac{r_2}{\log q}.
\end{align}
Also, we introduce 
the parameter $r_1$ for the sacrifice rate 
as
\begin{align}
\lim_{n\to \infty}\frac{\bar{k}_n}{n}
&= \frac{r_1}{\log q} .
\end{align}
Since $N(\vec{\lambda},G) \le  {n \choose \vec{\lambda}} $,
$r_2$ is bounded as 
 \begin{align}
 r_2 \le (\log q)\lim_{n\to \infty} \frac{n-k_n}{n}
 =\log q - r_0.
\Label{ALO}
\end{align}

\begin{table}[htpb]
  \caption{Summary of rates}
\label{T1}
\begin{center}
{
\renewcommand\arraystretch{1.7}
  \begin{tabular}{|c|l|l|} 
\hline
$r_0 $ & Rate of error correcting code&$\lim_{n\to \infty}\frac{k_n}{n} \log q $\\
\hline
$r_1$ & Sacrifice rate &$\lim_{n\to \infty}\frac{\bar{k}_n}{n} \log q $\\
\hline
$r_2 $ & Rate of $A$ & $\lim_{n\to \infty}\frac{\log A}{n} \log q $\\
\hline
  \end{tabular}}
\end{center}
\end{table}

Then, \eqref{LLLA} implies
\begin{align}
\lim_{n\to \infty}\frac{-1}{n}\log 
(\min_{s \in [0,\frac{1}{2}]} B_{i,n,s,1})
\ge \max_{s \in [0,\frac{1}{2}]}
s(r_1+r_0
-\log q- I_{\frac{1}{1-s}}^{\downarrow}(Y; X_1  )[W]),
\Label{HTRE}
\end{align}
and \eqref{LLL} implies
\begin{align}
\lim_{n\to \infty}\frac{-1}{n}\log 
(\min_{s \in [0,\frac{1}{2}]} B_{i,n,s,2} [A])
\ge 
&\max_{s \in [0,\frac{1}{2}]}
\min(
s(r_0+r_1- I_{\frac{1}{1-s}}^{\downarrow}(Y; X_1 ,X_2  )[W]),\nonumber \\
&
s(r_1-r_2- I_{\frac{1}{1-s}}^{\downarrow}(Y; X_1  )[W])).
\Label{HTYE}
\end{align}
Thus, the condition for the exponential decay of 
the upper bound $\min_{s \in [0,\frac{1}{2}]} B_{i,n,s,1}$
is the condition $r_1 > \log q +I(Y; X_1  )[W]- r_0 $.
That is,
when we use the upper bound $\min_{s \in [0,\frac{1}{2}]} B_{i,n,s,1}$
and the rate of error correcting code is fixed to $r_0$,
the achievable rate is the following value 
\begin{align}
 r_0-(\log q +I(Y; X_1  )[W]- r_0)=
2 r_0-\log q -I(Y; X_1  )[W], \Label{NFR}
\end{align}
which is called the 1st type of rate.
Also, the condition for the exponential decay of 
$\min_{s \in [0,\frac{1}{2}]} B_{i,n,s,2}[A]$ is 
the condition
$r_1 > \max(I(Y; X_1,X_2  )[W]- r_0,I(Y; X_1  )[W]+r_2 )$.
When we use the upper bound $\min_{s \in [0,\frac{1}{2}]} B_{i,n,s,1}$
and the rate of error correcting code is fixed to $r_0$,
the achievable rate is the following value 
\begin{align}
r_0-\max(I(Y; X_1,X_2  )[W]- r_0,I(Y; X_1  )[W]+r_2 )=
\min(
2 r_0-I(Y; X_1,X_2  )[W],r_0-r_2-I(Y; X_1  )[W]),
\Label{NFR2}
\end{align}
which is called the 2nd type of rate.
Since it is not so easy to calculate $r_2$ in the general case, 
we have the following lower bound of \eqref{NFR2} by substituting $r_2=0$; 
\begin{align}
\min(
2 r_0-I(Y; X_1,X_2  )[W],r_0-I(Y; X_1  )[W]),
\Label{Achi}
\end{align}
which is called the 3rd type of rate.
In fact, when these rates are negative, the achievable rates are zero.
However, to see the mathematical behaviors of the above differences, we address these values directly in this paper.

\section{Examples}\Label{S5}

\begin{figure}[t]
\begin{center}
\includegraphics[scale=0.9]{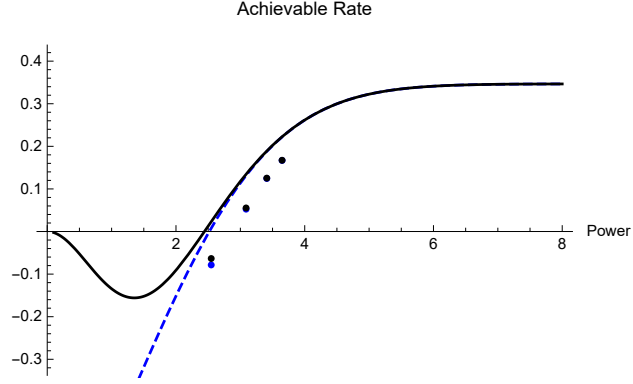}
\end{center}
\caption{Achievable rates with BPSK when the variance $N_0$ is $=1$.
The base of logarithm is chosen to be $e$.
The horizontal axis expresses the intensity $h$. 
The vertical axis expresses transmission rate.
The solid black line express 
the 2nd type of rate with random coding given as \eqref{H13}.
This value is positive with $h \ge 2.443$ and approaches $\frac{1}{2}\log 2$.
The dashed blue line express 
the 1st type of rate with random coding given as \eqref{H17}.
This value is positive with $h \ge 2.518$ and approaches $\frac{1}{2}\log 2$.
The black points express the 3rd type of rate with 
$(d_l,d_r,L)$ spatial coupling LDPC code with sufficiently large $L$,
whose rate is \eqref{H14}.
The blue points express the 1st type of rate with 
$(d_l,d_r,L)$ spatial coupling LDPC code with sufficiently large $L$,
whose rate is \eqref{H18}.
According to these formulas,
the value is negative when $h$ is less than a certain threshold.
In this case, the secure transmission of $M_1+M_2$ is impossible in these methods.}
\Label{F1}
\end{figure}%

\subsection{Random coding with universal$2$ condition}\Label{SeR}
For simplicity, we ignore the decoding time, and 
discuss the asymptotic transmission rate.
Then, 
the generating matrix $G \in \FF_q^{n \times k_n}$
is assumed to be generated subject to the universal$2$ condition \eqref{GDE}.
We employ the channel decoding for the degraded channel \cite{Ullah2}.
That is, the decoder is given as
\begin{align}
\argmax_{\bv \in \FF_q^{k_n}}
\sum_{i=1}^n \log \hat{\phi}_{h (G(\bv)_i-e_1-e_2),N_0}(Y_i),
\end{align}
where 
$\hat{\phi}_{x,N_0}(y):=\sum_{x'\in \FF_q} \frac{1}{q}{\phi}_{
h_1 \sigma(x')+h_2 \sigma(x-x'),N_0}(y)$.
Then, the optimal $k_n$ satisfies \cite{Ullah}
\begin{align}
r_0=\lim_{n\to \infty} \frac{k_n}{n}\log q
=&
I(Y;X_1+X_2)[W].
\end{align}
Then, the 1st type of rate is 
\begin{align}
2 r_0-\log q -I(Y; X_1  )[W]
=2 I(Y;X_1+X_2)[W]-\log q -I(Y; X_1  )[W].\Label{former}
\end{align}
Since $r_2=0$, 
the 2nd type of rate equals
the 3rd type of rate, which is calculated as
\begin{align}
2 I(Y;X_1+X_2)[W] -I(Y; X_1,X_2  )[W].
\Label{LODT}
\end{align}
See Appendix \ref{AC}.
Next, We consider the BPSK scheme, and assume that $h_1=h_2=h$. 
Then, $\hat{\phi}_{x,N_0}$ is simplified as 
$\hat{\phi}_{0,N_0}(y)=
({\phi}_{0,N_0}( y)+ {\phi}_{2h,N_0}( y ))/2$
and
$\hat{\phi}_{1,N_0}(y)={\phi}_{h,N_0}( y)$.
Then, 
by using the differential entropy $H$,
$r_0$ is calculated to be $I(h):=
H(\frac{\phi_{0,N_0}+2 \phi_{h,N_0}+\phi_{2h,N_0}}{4})
-\frac{1}{2} H(\frac{\phi_{0,N_0}+\phi_{2h,N_0}}{2})-\frac{1}{2} H(\phi_{h,N_0})$,
and the 2nd type of rate \eqref{Achi} is 
\begin{align}
2I(Y;X_1+X_2)[W] -I(Y; X_1 ,X_2  )[W]
=& H(\frac{\phi_{0,N_0}+2 \phi_{h,N_0}+\phi_{2h,N_0}}{4}) \nonumber \\
&- H(\frac{\phi_{0,N_0}+\phi_{2h,N_0}}{2}).\Label{H13}
\end{align}
In this case, since
$I(Y; X_1  )[W]=
H(\frac{\phi_{0,N_0}+2 \phi_{h,N_0}+\phi_{2h,N_0}}{4})- 
H(\frac{\phi_{0,N_0}+\phi_{2h,N_0}}{2})$,
due to \eqref{former},
the 1st type of rate \eqref{NFR} is 
\begin{align}
H(\frac{\phi_{0,N_0}+2 \phi_{h,N_0}+\phi_{2h,N_0}}{4})
- H(\phi_{h,N_0})- \log 2.\Label{H17}
\end{align}
That is, the difference between the 1st type and 2nd type of rates
are the value 
$\log 2 -
( H(\frac{\phi_{0,N_0}+\phi_{2h,N_0}}{2})- H(\phi_{h,N_0}))$.
This value becomes very small when $h$ is sufficiently large in comparison
with $N_0$ as Fig. \ref{F1}
because the two distributions $\phi_{0,N_0}$ and $\phi_{2h,N_0}$ can be distinguished with high probability.

\subsection{LDPC code}\Label{SeL}

In fact, it is not so easy to calculate the coefficient $A$ in a real code, e.g., an LDPC code.
In this case, we employ the finite-length security formula \eqref{LLLA} of Theorem \ref{TTA}.
Hence, we focus on the 1st type of rate \eqref{NFR} as the asymptotic transmission rate with the security guarantee of the finite-length setting.
Since it is not so easy calculate the 2nd type of rate,
we address the 3rd type of rate as its lower bound. 
When the difference between the 1st and the 3rd types of rates is small, 
we can conclude that the 2nd type of rate is close to the 1st type of rate.


Now, we consider the BPSK case with $h_1=h_2=h$ when $G$ is a $(d_l,d_r,L)$ spatial coupling LDPC code with large $L$.
According to the preceding papers \cite{KRU,Takabe}, we employ belief propagation in decoder.
Applying density evolution to the channel $x \mapsto \hat{\phi}_{x,N_0}$, 
the paper  \cite{Takabe} calculated the transmission rate $I_{sc}(h)$ in the code.
By using the difference $\Delta I(h):=I(h)-I_{sc}(h)$, 
the 3rd type of rate \eqref{Achi} is calculated to
\begin{align}
\lim_{n\to \infty} \frac{k_n-\bar{k}_n}{n}\log 2 
=& H(\frac{\phi_{0,N_0}+2 \phi_{h,N_0}+\phi_{2h,N_0}}{4}) \nonumber \\
&- H(\frac{\phi_{0,N_0}+\phi_{2h,N_0}}{2})-2 \Delta I(h).\Label{H14}
\end{align}
The 1st type of rate \eqref{NFR} is 
\begin{align}
&2 I(Y;X_1+X_2)[W]-\log 2 -I(Y; X_1  )[W]-2 \Delta I(h)\nonumber\\
=&
H(\frac{\phi_{0,N_0}+2 \phi_{h,N_0}+\phi_{2h,N_0}}{4})
- H(\phi_{h,N_0})- \log 2 -2 \Delta I(h).\Label{H18}
\end{align}
Fig. \ref{F1} shows that the difference between the 1st and the 3rd types of rates is small.

\section{Conclusion}\Label{S8}
In order to make secure transmission via untrusted relay,
we have derived a code that securely transmits the XOR of the messages of two nodes
via a multiple access channel.
In this code, the relay cannot obtain any information for the message of each node,
and can decode only the messages of two nodes.
Since our code is constructed by simple combination of an existing liner code and universal2 hash function,
it can be realizable in practice. 

To apply this system to a real secure satellite communication,
we need to study the following items.
First, we need to evaluate the performance of the proposed LDPC codes 
with a finite-length setting, which requires computer simulation.
Then, it is needed to combine the result of this computer simulation and
the security evaluation based on \eqref{LLLA} of Theorem \ref{TTA}.

Further, we need to consider the case when the relay does not inform the correct values of the strength of $h$ and $N_0$ to both nodes.
That is, there is a possibility that
the true values of $h/N_0$ is larger than the value informed by the relay.
In this case, both nodes can estimate the upper bound of $h/N_0$
by using the spatial conditions.
Then, for evaluation of the decoding error probability, 
both nodes need to use the value of $h/N_0$ informed by the relay.
For evaluation of the amount of leaked information, 
both nodes need to use the upper bound of $h/N_0$.
For real implementation, it is needed to numerically simulate 
the security evaluation based on this observation.

Finally, we should remark that Theorems \ref{TTA} and \ref{TTY} cannot be shown by simple application of the result of wire-tap channel \cite{Wyner} as follows.
Consider the secrecy of the message $M_1$ of node $1$.
In this case, if node $2$ transmits elements of $\FF_q^n$ with equal probability,
the channel from node $1$ to relay $R$ 
is given as $n$-fold extension of the degraded channel 
$x \mapsto \sum_{x'\in \FF_q}\frac{1}{q} \phi_{h_1 \sigma(x)+ h_2 \sigma(x-x'),N_0}$, which enables us to directly apply the result of wire-tap channel.
However, node $2$ transmits elements of the image of  $G$, which is a subset of $\FF_q^n$, with equal probability.
Hence, the channel from node $1$ to relay $R$ does not have the above simple form.
Therefore, we need more careful discussion.
Finally, we point out that
or proofs of Theorems \ref{TTA} and \ref{TTY} are still valid even when 
the channel is a general multiple access channel whose input is given as
$\FF_q \times \FF_q $
because our proofs employ only the property of a general multiple access channel.

\section*{Acknowledgments}
We are grateful to Dr. Satoshi Takabe for giving the numerical values for Fig. \ref{F1}.
The work reported here was supported in part by 
the JSPS Grant-in-Aid for Scientific Research 
(A) No.17H01280, (B) No. 16KT0017, (C) No. 16K00014, 
and Kayamori Foundation of Informational Science Advancement.

\appendices

\section{Proof of Lemma \ref{GUR}}\Label{AA}
For the proof of Lemma \ref{GUR}, we prepare the following lemma.
\begin{lemma}\Label{LOE}
When the conditional distribution $P_{Z_2|Z_1}$ is the uniform distribution on ${\cal Z}_2$, we have
\begin{align}
e^{s I_{\frac{1}{1-s} }^{\downarrow}( Y;Z_1,Z_2)}
\le 
|{\cal Z}_2|^s
e^{s I_{\frac{1}{1-s} }^{\downarrow}( Y;Z_1)}
\Label{HY2}
\end{align}
for $s \in [0, \infty)$.
\end{lemma}

\begin{proofof}{Lemma \ref{LOE}}
We have
\begin{align*}
& e^{s I_{\frac{1}{1-s} }^{\downarrow}( Y;Z_1,Z_2)}
=
\int_{{\cal Y}}  
\Big(
\sum_{z_1}
P_{Z_1}(z_1)
\sum_{z_2}
P_{Z_2|Z_1}(z_2|z_1)
p_{Y|Z_1,Z_2}(y|z_1,z_2)^{\frac{1}{1-s}}
\Big)^{1-s}
dy \\
= &
\int_{{\cal Y}}  
\Big(
\sum_{z_1}
P_{Z_1}(z_1)
|{\cal Z}_2|^{\frac{s}{1-s}}
\sum_{z_2}
P_{Z_2|Z_1}(z_2|z_1)^{\frac{1}{1-s}} 
p_{Y|Z_1,Z_2}(y|z_1,z_2)^{\frac{1}{1-s}}
\Big)^{1-s}
dy \\
\le &
\int_{{\cal Y}}  
\Big(
\sum_{z_1}
P_{Z_1}(z_1)
|{\cal Z}_2|^{\frac{s}{1-s}}
\Big(
\sum_{z_2}
P_{Z_2|Z_1}(z_2|z_1)
p_{Y|Z_1,Z_2}(y|z_1,z_2)
\Big)^{\frac{1}{1-s}}
\Big)^{1-s}
dy \\
= &
|{\cal Z}_2|^{s}
\int_{{\cal Y}}  
\Big(
\sum_{z_1}
P_{Z_1}(z_1)
p_{Y|Z_1}(y|z_1)^{\frac{1}{1-s}}
\Big)^{1-s}
dy \\
=&
|{\cal Z}_2|^s
e^{s I_{\frac{1}{1-s} }^{\downarrow}( Y;Z_1)}.
\end{align*}
\end{proofof}

\begin{proofof}{Lemma \ref{GUR}}
Since $N(\vec{\lambda},G) \le  {n \choose \vec{\lambda}} $,
$A$ is bounded as $q^{n-k_n}$.
Hence, 
\begin{align*}
 3 q^{s (n-k_n-\bar{k}_n)} e^{s n I_{\frac{1}{1-s}}^{\downarrow}
(Y; X_1  )[W] }
\ge 
3 A^s q^{-s \bar{k}_n} e^{s n I_{\frac{1}{1-s}}^{\downarrow}
(Y; X_1  )[W] }.
\end{align*}
Also, Lemma \ref{LOE} guarantee that 
$ 3 q^{s (n-k_n-\bar{k}_n)} e^{s n I_{\frac{1}{1-s}}^{\downarrow}
(Y; X_1  )[W] }  $ is less than the first term of RHS of \eqref{LLL7}.
Hence, we obtain \eqref{LLL6}.
\end{proofof}

\section{Proof of Theorems \ref{TTA} and \ref{TTY}}\Label{S7}
\noindent{\bf Step (1):}\quad 
First, we notice the relation
\begin{align}
\bV_i= F_1(M_i)+F_2(L_i), \quad 
\bX_i= (\bV_i,E_i).
\end{align}
Since $(E_1,E_2)$ is subject to the uniform distribution on 
$ \FF_q^{2(n -k_n)}$, 
even when the map $G$ is fixed to be $g$,
$(\bX_1,\bX_2)$ is subject to the uniform distribution on 
$ {\cal X}^{2n}$.
Receiver receives the random variable $\bY \in {\cal Y}^n$ that  depends only on $(\bX_1,\bX_2)$.
Once, $G$ is fixed, we have the Markov chain $(F, M_1,M_2,L_1,l_2,E_1,E_2)- (\bX_1,\bX_2) -\bY$.
Due to \eqref{23-1}, the relay node can decode $M_1+M_2$ and $L_1+l_2$ 
from $\bY$ by using the knowledge $E_1+E_2$
for the coset.

Then, we focus on the randomness of the choice of $F$. 
Then, for $s \in [0,\frac{1}{2}]$, we have
\begin{align}
& \EE_{E_1, E_2,F}
\| P_{M_1 \bY| E_1, E_2 ,n}-  
P_{\bY| E_1, E_2 ,n}\times P_{M_1,\mix}\|_1  \nonumber \\
\stackrel{(a)}{\le} & 
3 q^{-s \bar{k}_n} e^{s I_{\frac{1}{1-s}}^{\downarrow}
(\bY; \bV_1 | G=g, E_1, E_2 ) } 
\nonumber\\
\stackrel{(b)}{\le} & 
3 q^{-s \bar{k}_n}
e^{s I_{\frac{1}{1-s}}^{\downarrow}
(\bY; \bV_1, E_1 | G=g, E_2 ) } 
\nonumber\\
\stackrel{(c)}{=} &  
3q^{-s \bar{k}_n}
e^{s I_{\frac{1}{1-s}}^{\downarrow}
(\bY; \bX_1 | G=g, E_2 ) } 
\Label{TY10} \\
\stackrel{(d)}{\le} & 
3 q^{s(n-k_n- \bar{k}_n)} 
 e^{s I_{\frac{1}{1-s}}^{\downarrow}(\bY; \bX_1  ) } 
\nonumber\\
= & 
3 q^{s(n-k_n- \bar{k}_n)} 
e^{sn I_{\frac{1}{1-s}}^{\downarrow}(Y; X_1  )[W] } ,
\Label{TUY}
\end{align}
where $(a)$ follows from Theorem 6 of \cite{Hayashi2013} and the universal2 condition for $F$;
$(b)$ follows from \eqref{HY2};
$(c)$ follows from the fact that the pair $(E_i,F_i)$ and $\bX_i$
uniquely determine each other;
and
$(d)$ will be shown in the next step.
Hence, we obtain \eqref{LLLA} in Theorem \ref{TTA}.

Now, we proceed to the proof of Theorem \ref{TTY}.
\begin{align}
&3q^{-s \bar{k}_n}
e^{s I_{\frac{1}{1-s}}^{\downarrow}
(\bY; \bX_1 | G, E_2 ) } 
\nonumber\\
\stackrel{(e)}{\le} & 
3q^{-s (k_n+\bar{k}_n)}
e^{s I_{\frac{1}{1-s}}^{\downarrow}(\bY; \bX_1 ,\bX_2  ) } 
+3A^{s} q^{-s \bar{k}_n} e^{s I_{\frac{1}{1-s}}^{\downarrow}(\bY; \bX_1  ) } 
\nonumber\\
= & 
3q^{-s (k_n+\bar{k}_n)}
e^{sn I_{\frac{1}{1-s}}^{\downarrow}(Y; X_1 ,X_2  )[W] } 
+3A^{s} q^{-s \bar{k}_n} 
e^{sn I_{\frac{1}{1-s}}^{\downarrow}(Y; X_1  )[W] } ,
\Label{TUY2}
\end{align}
where $(e)$ will be shown in the next step.
Combing \eqref{TY10} and \eqref{TUY2},
we obtain \eqref{LLL} in Theorem \ref{TTY}.

\noindent{\bf Step (2):}\quad 
Now, we show $(d)$ in \eqref{TUY}.
Given a code $g:\FF_q^n\to \FF_q^{k_n}$, 
for an element $\bx \in \FF_q^n$, 
we uniquely have a coset $[\bx]$ and its representative $e \in \FF_q^n$.
We denote the map from an element $\bx \in \FF_q^n$ 
to the representative $e \in \FF_q^n$ by $g_2$.
We define the set 
${\cal S}(g_2,e):=\{ \bx \in \FF_q^n | g_2(\bx_j)=e\}$. 
Definition of $A$ implies
\begin{align}
&
\sum_{ \bx_2' \in {\cal S}(g_2,g_2(\bx_2)) }
P_{Y|\bX_1=\bx_1 ,\bX_2=\bx_2'} (\by)
\nonumber \\
\le &
\sum_{ \bx_2'}
P_{Y|\bX_1=\bx_1 ,\bX_2=\bx_2'} (\by)
\nonumber \\
=& q^{n}P_{Y|\bX_1=\bx_1 } (\by).\Label{HY12B}
\end{align}

Using \eqref{HY12}, for $s \in [0,\frac{1}{2}]$,
we have the following relations, 
where the explanations for steps is explained later.
\begin{align}
& e^{s I_{\frac{1}{1-s}}^{\downarrow}
(\bY; \bX_1 | G=g, E_2 ) } 
\nonumber\\
= &
\EE_{E_2}
\int_{{\cal Y}^n} \Big(
q^{-n} \sum_{\bx_1}
P_{Y|\bX_1=\bx_1 , E_2 ,G=g}(\by)^{\frac{1}{1-s}}
\Big)^{1-s}
d \by 
\nonumber \\
= &
\EE_{E_2}
\int_{{\cal Y}^n} 
\Big(
q^{-n}\sum_{\bx_1}
\Big(
q^{-k_n}
\sum_{ \bx_2 \in {\cal S}(g_2,E_2) }
P_{Y|\bX_1=\bx_1 ,\bX_2=\bx_2}
 (\by)
\Big)^{\frac{1}{1-s}}
\Big)^{1-s}
d \by 
\nonumber \\
\stackrel{(a)}{\le} & 
\int_{{\cal Y}^n}
\Big(
 q^{-n}\sum_{\bx_1}
\EE_{ E_2}
\Big(
q^{-k_n}
\sum_{ \bx_2 \in {\cal S}(g_2,E_2) }
P_{Y|\bX_1=\bx_1 ,\bX_2=\bx_2}
 (\by)
\Big)^{\frac{1}{1-s}}
\Big)^{1-s}
d \by 
\nonumber \\
= & 
\int_{{\cal Y}^n} 
\Big(
q^{-n}\sum_{\bx_1}
\EE_{E_2}
q^{-\frac{k_n}{1-s}}
\Big(
\sum_{ \bx_2 \in {\cal S}(g_2,E_2) }
P_{Y|\bX_1=\bx_1 ,\bX_2=\bx_2}
 (\by)
\Big(
\nonumber \\
&\sum_{ \bx_2' \in {\cal S}(g_2,E_2) }
P_{Y|\bX_1=\bx_1 ,\bX_2=\bx_2'} (\by)
\Big)^{\frac{s}{1-s}}
\Big)
\Big)^{1-s}d \by 
\nonumber \\
= & 
\int_{{\cal Y}^n} \Big(
q^{-n}\sum_{\bx_1}
q^{-\frac{k_n}{1-s}}
\Big(
q^{k_n-n} \sum_{ \bx_2 }
P_{Y|\bX_1=\bx_1 ,\bX_2=\bx_2} (\by)
\Big(
\sum_{ \bx_2' \in {\cal S}(g_2,g_2(\bx_2)) }
P_{Y|\bX_1=\bx_1 ,\bX_2=\bx_2'} (\by)
\Big)^{\frac{s}{1-s}}
\Big)
\Big)^{1-s}d \by 
\nonumber \\
\stackrel{(b)}{\le} & 
\Big(
\int_{{\cal Y}^n} q^{-n}\sum_{\bx_1}
q^{-\frac{k_n}{1-s}}
\Big(
q^{k_n-n} \sum_{ \bx_2 }
P_{Y|\bX_1=\bx_1 ,\bX_2=\bx_2} (\by)
\Big(
q^{n}
P_{Y|\bX_1=\bx_1 } (\by)
\Big)^{\frac{s}{1-s}}
\Big)
\Big)^{1-s} d \by 
\nonumber \\
= & 
\int_{{\cal Y}^n} \Big(
q^{-n}\sum_{\bx_1}
q^{-\frac{k_n}{1-s}}
\Big(
q^{k_n} 
P_{Y|\bX_1=\bx_1 } (\by)
\Big(
q^{n}
P_{Y|\bX_1=\bx_1 } (\by)
\Big)^{\frac{s}{1-s}}
\Big)
\Big)^{1-s}
d \by \nonumber \\
= & 
q^{s (n-k_n)}
\int_{{\cal Y}^n} \Big(
q^{-n}\sum_{\bx_1}
P_{Y|\bX_1=\bx_1 } (\by)^{\frac{1}{1-s}}
\Big)^{1-s}d \by 
\nonumber \\
=&
q^{s (n-k_n)}
 e^{s I_{\frac{1}{1-s}}^{\downarrow}
(\bY; \bX_1  ) } ,
\end{align}
where 
$(a)$ follows from the concavity of $x \mapsto x^{1-s}$ with $x\ge 0$.
and
$(b)$ follows from the concavity of \eqref{HY12B}.

\noindent{\bf Step (3):}\quad 
Now, we show $(e)$ in \eqref{TUY2}.
Definition of $A$ implies
\begin{align}
&\EE_{G}
\sum_{ \bx_2' (\neq \bx_2)\in {\cal S}(G_2,G_2(\bx_2)) }
P_{Y|\bX_1=\bx_1 ,\bX_2=\bx_2'} (\by)
\nonumber \\
=&
\sum_{\vec{\lambda}\neq \vec{0}_n}
\sum_{ \bx_2': \vec{n}(\bx_2'- \bx_2)= \vec{\lambda}}
P( \bx_2'- \bx_2 \in \im G)
P_{Y|\bX_1=\bx_1 ,\bX_2=\bx_2'} (\by)
\nonumber \\
=&
\sum_{\vec{\lambda}\neq \vec{0}_n}
\sum_{ \bx_2': \vec{n}(\bx_2'- \bx_2)= \vec{\lambda}}
\EE_{G} \frac{N(\vec{\lambda},G)}{ {n \choose \vec{\lambda}} }
P_{Y|\bX_1=\bx_1 ,\bX_2=\bx_2'} (\by)
\nonumber \\
\le &
A q^{k_n-n}
\sum_{\vec{\lambda}\neq \vec{0}_n}
\sum_{ \bx_2': \vec{n}(\bx_2'- \bx_2)= \vec{\lambda}}
P_{Y|\bX_1=\bx_1 ,\bX_2=\bx_2'} (\by)
\nonumber \\
\le & 
A q^{k_n}
P_{Y|\bX_1=\bx_1} (\by).\Label{HY12}
\end{align}

Using \eqref{HY12}, for $s \in [0,\frac{1}{2}]$,
we have the following relations, 
where the explanations for steps is explained later.
\begin{align}
& e^{s I_{\frac{1}{1-s}}^{\downarrow}
(\bY; \bX_1 | G, E_2 ) } 
\nonumber\\
= &
\EE_{G, E_2}
\int_{{\cal Y}^n}\Big(
 q^{-n}\sum_{\bx_1}
P_{Y|\bX_1=\bx_1 , E_2 ,G}(\by)^{\frac{1}{1-s}}
\Big)^{1-s} d \by 
\nonumber \\
= &
\EE_{G, E_2}
\int_{{\cal Y}^n} \Big(
q^{-n}\sum_{\bx_1}
\Big(
q^{-k_n}
\sum_{ \bx_2 \in {\cal S}(G_2,E_2) }
P_{Y|\bX_1=\bx_1 ,\bX_2=\bx_2}
 (\by)
\Big)^{\frac{1}{1-s}}
\Big)^{1-s}d \by 
\nonumber \\
\stackrel{(a)}{\le} & 
\int_{{\cal Y}^n} \Big(
q^{-n}\sum_{\bx_1}
\EE_{G, E_2}
\Big(
q^{-k_n}
\sum_{ \bx_2 \in {\cal S}(G_2,E_2) }
P_{Y|\bX_1=\bx_1 ,\bX_2=\bx_2}
 (\by)
\Big)^{\frac{1}{1-s}}
\Big)^{1-s}d \by 
\nonumber \\
= & 
\int_{{\cal Y}^n} \Big(
q^{-n}\sum_{\bx_1}
\EE_{G,E_2}
q^{-\frac{k_n}{1-s}}
\Big(
\sum_{ \bx_2 \in {\cal S}(G_2,E_2) }
P_{Y|\bX_1=\bx_1 ,\bX_2=\bx_2}
 (\by)
\Big(
P_{Y|\bX_1=\bx_1 ,\bX_2=\bx_2} (\by)
\nonumber \\
&+\sum_{ \bx_2' (\neq \bx_2)\in {\cal S}(G_2,E_2) }
P_{Y|\bX_1=\bx_1 ,\bX_2=\bx_2'} (\by)
\Big)^{\frac{s}{1-s}}
\Big)
\Big)^{1-s}d \by 
\nonumber \\
= & 
\int_{{\cal Y}^n} \Big(
q^{-n}\sum_{\bx_1}
\EE_{G}
q^{-\frac{k_n}{1-s}}
\Big(
q^{k_n-n} \sum_{ \bx_2 }
P_{Y|\bX_1=\bx_1 ,\bX_2=\bx_2} (\by)
\Big(
P_{Y|\bX_1=\bx_1 ,\bX_2=\bx_2} (\by)
\nonumber \\
&+\sum_{ \bx_2' (\neq \bx_2)\in {\cal S}(G_2,G_2(\bx_2)) }
P_{Y|\bX_1=\bx_1 ,\bX_2=\bx_2'} (\by)
\Big)^{\frac{s}{1-s}}
\Big)
\Big)^{1-s}d \by 
\Label{GTR} \\
\stackrel{(b)}{\le} & 
\int_{{\cal Y}^n} \Big(
q^{-n}\sum_{\bx_1}
q^{-\frac{k_n}{1-s}}
\Big(
q^{k_n-n} \sum_{ \bx_2 }
P_{Y|\bX_1=\bx_1 ,\bX_2=\bx_2} (\by)
\Big(
P_{Y|\bX_1=\bx_1 ,\bX_2=\bx_2} (\by)^{\frac{s}{1-s}}
\nonumber \\
&+
\Big(\EE_{G}
\sum_{ \bx_2' (\neq \bx_2)\in {\cal S}(G_2,G_2(\bx_2)) }
P_{Y|\bX_1=\bx_1 ,\bX_2=\bx_2'} (\by)
\Big)^{\frac{s}{1-s}}
\Big)
\Big)
\Big)^{1-s}d \by 
\nonumber \\
\stackrel{(c)}{\le} & 
\int_{{\cal Y}^n} \Big(
q^{-n}\sum_{\bx_1}
q^{-\frac{k_n}{1-s}}
\Big(
q^{k_n-n} \sum_{ \bx_2 }
P_{Y|\bX_1=\bx_1 ,\bX_2=\bx_2}(\by)
\Big(
P_{Y|\bX_1=\bx_1 ,\bX_2=\bx_2} (\by)^{\frac{s}{1-s}}
\nonumber \\
&+
\Big(
A q^{k_n}
P_{Y|\bX_1=\bx_1} (\by)
\Big)^{\frac{s}{1-s}}
\Big)
\Big)
\Big)^{1-s}d \by 
\nonumber \\
= & 
\int_{{\cal Y}^n} \Big(
q^{-n}\sum_{\bx_1}
q^{-\frac{k_n}{1-s}}
\Big(
q^{k_n-n} \sum_{ \bx_2 }
P_{Y|\bX_1=\bx_1 ,\bX_2=\bx_2}(\by)^{\frac{1}{1-s}} 
+
A^{\frac{s}{1-s}} q^{\frac{sk_n }{1-s}+k_n}
P_{Y|\bX_1=\bx_1} (\by)^{\frac{1}{1-s}}
\Big)
\Big)^{1-s}d \by 
\nonumber \\
= & 
\int_{{\cal Y}^n}
\Big(
q^{-\frac{s k_n}{1-s}}
q^{-2 n} \sum_{ \bx_1,\bx_2 }
P_{Y|\bX_1=\bx_1 ,\bX_2=\bx_2}(\by)^{\frac{1}{1-s}} 
+ A^{\frac{s}{1-s}}
q^{-n}\sum_{\bx_1}
P_{Y|\bX_1=\bx_1} (\by)^{\frac{1}{1-s}}
\Big)^{1-s}
d \by \nonumber \\
\stackrel{(d)}{\le} & 
\int_{{\cal Y}^n}
\Big(
q^{-\frac{s k_n}{1-s}}
q^{-2 n} \sum_{ \bx_1,\bx_2 }
P_{Y|\bX_1=\bx_1 ,\bX_2=\bx_2} (\by)^{\frac{1}{1-s}}
\Big)^{1-s} d \by 
+ 
\int_{{\cal Y}^n}\Big(
A^{\frac{s}{1-s}}
q^{-n}\sum_{\bx_1}
P_{Y|\bX_1=\bx_1} (\by)^{\frac{1}{1-s}}
\Big)^{1-s}d \by 
\nonumber \\
= & 
q^{-s k_n}
\int_{{\cal Y}^n}
\Big(
q^{-2 n} \sum_{ \bx_1,\bx_2 }
P_{Y|\bX_1=\bx_1 ,\bX_2=\bx_2}(\by)^{\frac{1}{1-s}} 
\Big)^{1-s}d \by 
+ A^{s}
\int_{{\cal Y}^n}
\Big(
q^{-n}\sum_{\bx_1}
P_{Y|\bX_1=\bx_1} (\by)^{\frac{1}{1-s}}
\Big)^{1-s}d \by 
\nonumber \\
=&
q^{-s k_n}
e^{s I_{\frac{1}{1-s}}^{\downarrow}(\bY; \bX_1 ,\bX_2  ) } 
+A^s e^{s I_{\frac{1}{1-s}}^{\downarrow}
(\bY; \bX_1  ) } ,
\end{align}
where each step can be shown as follows.
$(a)$ follows from the concavity of $x \mapsto x^{1-s}$ with $x\ge 0$.
$(b)$ follows from the concavity of $x \mapsto x^{\frac{s}{1-s}}$ with $x\ge 0$.
$(c)$ follows from the inequality \eqref{HY12}. 
$(d)$ follows from the inequality 
$(x+y)^{\frac{s}{1-s}}\le 
x^{\frac{s}{1-s}}+y^{\frac{s}{1-s}}$ with $x,y\ge 0$.

\section{Proof of \eqref{LODT}}\Label{AC}
\begin{align}
&\min(2 r_0-I(Y; X_1,X_2  )[W],r_0-I(Y; X_1  )[W]) \nonumber \\
=&
\min(2 I(Y;X_1+X_2)[W] -I(Y; X_1,X_2  )[W],
I(Y;X_1+X_2)[W] -I(Y; X_1  )[W])\nonumber \\
=&
2 I(Y;X_1+X_2)[W] -I(Y; X_1,X_2  )[W],
\Label{LOD}
\end{align}
where \eqref{LOD} is shown as follows.
Since $X_1+X_2$ and $X_1$ are independent of each other,
we have $I(Y; X_1 ,X_2  )[W]-I(Y;X_1+X_2)[W]
=I(Y; X_1 | X_1+X_2 )[W] $.
Therefore, 
\begin{align}
& I(Y;X_1+X_2)[W] -I(Y; X_1  )[W]
\ge I(Y;X_1+X_2)[W] -I(Y; X_1 | X_1+X_2 )[W]  \nonumber \\
=& 2I(Y;X_1+X_2)[W] -I(Y; X_1 ,X_2  )[W],
\end{align}
which implies \eqref{LOD}.





\end{document}